# Vacuum Quality Monitoring of Analytical Sputter Depth Profile Equipment by Recontamination Measurements of a Sputtered Ti Surface

Uwe Scheithauer, 82008 Unterhaching, Germany
Phone: + 49 89 636 – 44143 , E-Mail: scht.uhg@googlemail.com

**Abstract:**

Often a surface sensitive analytical technique in combination with sample erosion by inert ion sputtering is used for compositional in-depth analysis of solid state samples. Layer by layer the sample gets eroded and then the composition of the actual surface is estimated. The elemental detection limits can be increased by spending more time for the measurement in each sputter depth of the profile measurement. But during this time a significant sample surface recontamination with the element under investigation via adsorption form the vacuum has to be avoided.

Commonly the vacuum quality of an analytical instrument is monitored using the base pressure of the UHV system. This article presents a novel method, which improves this crude approach. The recontamination of a sputtered Ti surface by adsorbed residual gas particles was used to monitor the sputter depth profile specific vacuum condition of an XPS microprobe Quantum 2000 over a 4 years period. This new method is suitable to monitor the condition of every sputter depth profiling instrument.

________________________________________

## 1. Introduction

Often composition analysis of thin film systems utilizes in-depth profiling of the samples as the method of choice. Especially, Auger electron spectroscopy and X-ray photoelectron spectroscopy (XPS) microprobes are used for this kind of measurements. Applying in-depth profiling first the sample surface is eroded by ion bombardment ("sputtering") usually using inert gas ions in the energy range between 250 eV and 5 keV. Then the residual surface is analyzed after each sputter erosion step. The depth distributions of the elements are recorded as a function of sputter time [1]. To be aware of misleading results concerning the thin film composition, a recontamination of the actual sputtered surface by adsorption from the residual vacuum during the measurement itself has to be as low as possible. For example: The O content in the layers and at the interfaces due to the deposition process is under investigation. The unavoidable recontamination from the vacuum defines the detection limits during the measurements in a certain sputter depth.

Commonly the vacuum quality of an analytical instrument is monitored by using the base pressure of the ultra high vacuum (UHV) system. Measurements of the recontamination of a sputtered surface can give much more precise information about the sputter depth profile relevant vacuum quality. This article presents a development of such a procedure, which uses the O recontamination of a sputtered Ti surface. The results shown here have been elaborated in a long time monitoring of Ti recontamination rates of an analytical XPS microprobe Quantum 2000.

## 2. Instrumentation

For the measurements presented here a Physical Electronics XPS Quantum 2000 was used. This XPS microprobe achieves its spatial resolution by the combination of a fine-focused electron beam generating the X-rays on a water cooled Al anode and an elliptical mirror quartz monochromator, which monochromatizes and refocuses the X-rays to the sample surface. Details of the instruments design and performance are discussed elsewhere [2-8].

For sputter depth profiling the instrument is equipped with a differentially pumped $Ar^+$ ion gun. Sputter ion energies between 250 eV and 5 keV can be selected. The sputter rates are given in nm $SiO_2$ / min.. Thermally grown $SiO_2$ on a Si wafer, whose thickness was estimated by ellipsometry, is used as reference material for sputter erosion rate calibration [9].

For a flat mounted sample as used here in a Quantum 2000 the incoming X-rays are parallel to the surface normal. In this geometrical situation, the mean geometrical energy analyzer take off axis and the differentially pumped $Ar^+$ ion gun are oriented ~ 45° relative to the sample surface normal.

The samples were mechanically mounted on a 75mm x 75mm sample holder. This sample holder is introduced into the XPS vacuum chamber via a turbo pumped intro chamber.

Data evaluation was done by the PHI software Multipak 6.1 [10]. In case of quantification of measured peak intensities it uses the simplifying model, that all detected elements are distributed homogeneously within the analyzed volume. This volume is defined by the analysis area and the information depth of an XPS measurement, which is derived from the mean free path of electrons [11]. Using this approach one monolayer on top of a sample quantifies to ~ 10 ... 30 at% depended on the samples details.

## 3. Experimental Results

Test samples are a pure polycrystalline Ti sheet and a 99.8% pure polycrystalline Cu foil (NPL reference metal samples SCAA90, 419).

Fig. 1 shows XPS survey spectra of Ti and Cu samples, respectively. The samples were sputtered by 2 keV $Ar^+$ ions before the measurement. In the upper part of fig. 1 a survey spectrum was taken directly after sputtering the Ti sample. It takes ~ 30 min. to measure the data. During this time the surface was contaminated from the vacuum system with C (1.3 at%) and O (5.1 at%). Most likely water and OH groups are the source of the O contamination of the Ti surface. In the analytical instrument a water background is unavoidable due to the sample introduction process via the turbo pumped sample intro. Therefore on the Ti surface a formation of an oxide/hydroxide layer is expected [12]. The Ar sputter gas was implanted into the Ti sample. The middle of fig. 1 shows a survey spectrum of a sputtered Ti sample, which was stored in the vacuum for a longer time. The contamination layer has grown up. More C (28 at%) and O (48.3 at%) are detected. Additionally N (0.9 at%), F (1 at%) and a little amount of Ag (< 0.1 at%) are present. Since during the storage time of the Ti sample other samples were measured using the XPS instrument, most likely these last-mentioned contaminations are due to even these measurements. At least concerning the Ag contamination a reflective sputtering process must have taken place via the instrument inner surfaces [13]. The lower part of fig. 1 shows a survey spectrum of a sputtered Cu sample. The spectrum was measured directly after sputtering. It takes ~ 46 min. to measure it. No surface contamination with C (detection limit: 0.9 at%) and O (detection limit: 0.3 at%) is detected.

The XPS measurements show a much higher recontamination of the sputtered Ti in comparison to the sputtered Cu. This is expected, because Ti has a high sticking coefficient for gas absorption [14]. Just because of this property Ti is used as active material in ion getter pumps and sublimation pumps.

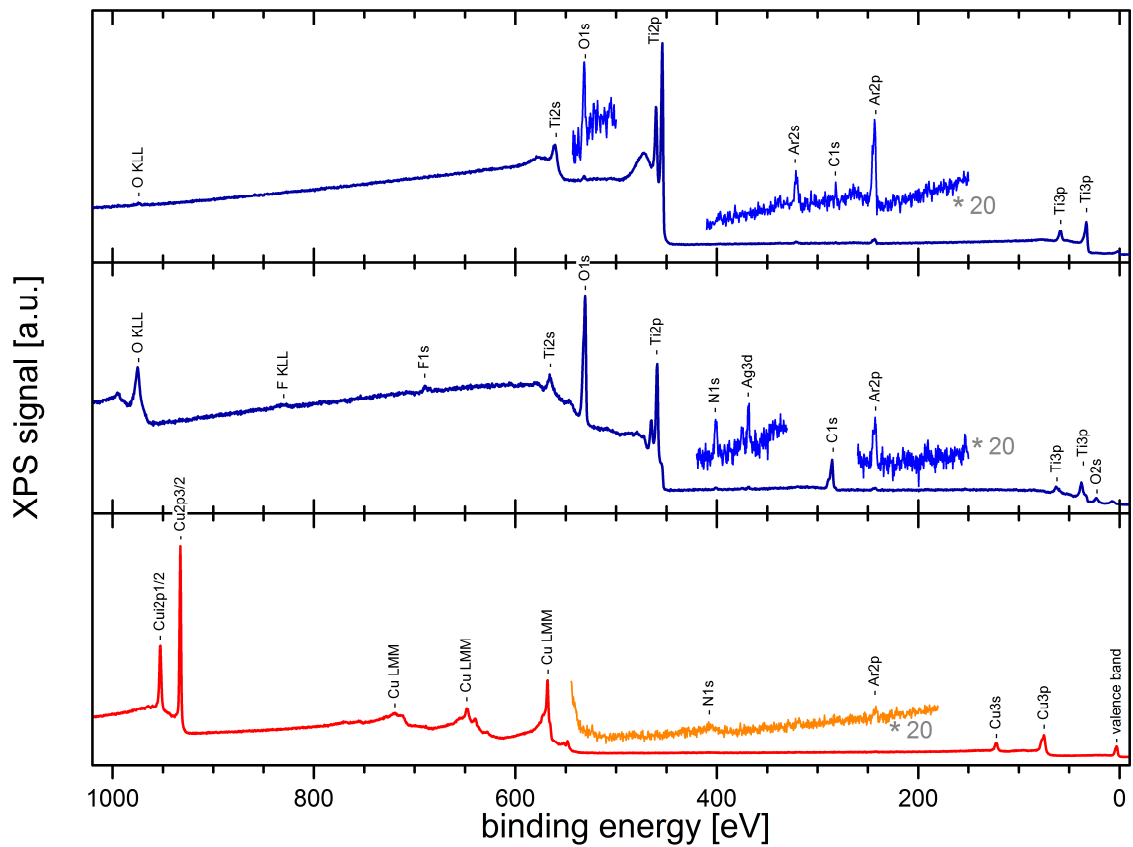

Fig. 1: recontamination of cleaned Ti and Cu samples
Top: Ti, $Ar^+$ sputtered, during the measurement time of ~ 30 min. the surface adsorbs contaminations, composition besides Ti: Ar (3.4 at%), C (1.3 at%), O (5.1 at%)
Middle: Ti, $Ar^+$ sputtered, stored in vacuum for longer time, composition besides Ti: Ar (0.6 at%), C (28 at%), O (48.3 at%), N (0.9 at%), F (1 at%), Ag (< 0.1 at%)
Bottom: Cu, $Ar^+$ sputtered, measurement time: ~ 46 min., composition besides Cu: Ar (1.1 at%), C (< 0.9 at%), O (< 0.3 at%)
C and O were not detected on the Cu surface, the detection limit is given.

The next experiments will show, that the vacuum quality can be characterized using the recontamination of Ti. The Ti is sputtered using 2 keV $Ar^+$ ions. With the applied ion dose 30 nm of $SiO_2$ could be removed. Approximately 250 nm of Ti are removed due to the higher sputter yield of Ti (sputter yield: ~ 8...10) in comparison to $SiO_2$ (sputter yield: ~ 1) [15, 16]. Then the ion gun remains in standby. Spectra of the O1s and Ti2p signal are measured with a repeating rate of 230 seconds. To obtain high count rates and thus a good O detection limit, the data were measured utilizing a high power X-ray beam (~ 200 μm diameter, ~ 45 W) and a low energy resolution of the energy analyzer. For the same reason ~ 85% of the 230 seconds measurement time is used to record the O1s signal.

Exemplarily a result of such a measurement is shown in fig. 2. Against the time elapsed after sputter cleaning of the Ti sample the O contamination of the surface is plotted. The quantification of the O signal relative to the Ti was calculated as described above. The O surface contamination increases linear with time for the first 3000 seconds. Then the curve flattens more and more. Corresponding to the results shown in fig.1, middle, the O signal should approach to a limit of ~50 at% after very long times. The gradient of the linear fit describes the recontamination rate of the sputter-cleaned Ti surface. At the beginning a constant O background of ~ 2.4 at% is detected. A recontamination during the first measurement interval, which lasts 230 sec., has to be taken into account. A part of the O signal is due to O at the grain boundaries of the polycrystalline Ti sheet. And additionally O may be implanted to the sample by the sputter process itself, if the ion gun is contaminated. Usually the ion gun contribution should decrease, if the ion gun was used frequently for longer time periods and the UHV was maintained properly. At minimum an O background of ~ 1.2 at% was detected during the Ti recontamination measurements, which were done over a period of some years.

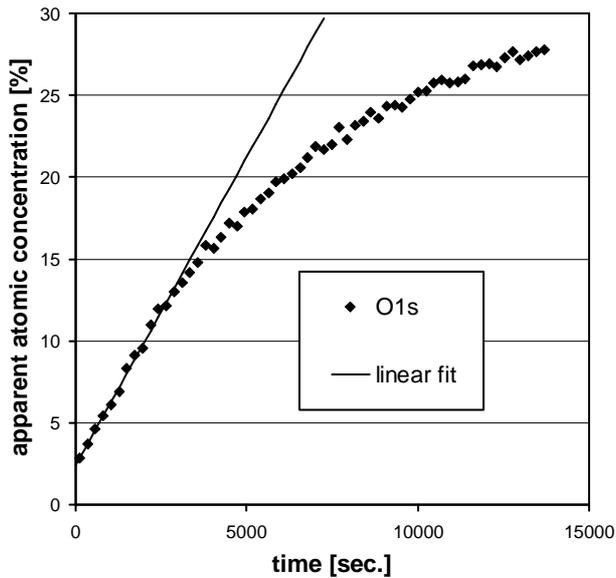

Fig. 2: O contamination of Ti as function of time elapsed after sputter cleaning

Fig. 3 summarizes the results of recontamination rate measurements over a 4 years period. Against the pressure of the UHV system the recontamination rate is plotted. As already mentioned, during each measurement the $Ar^+$ ion gun is in a standby mode. In this mode the Ar supply is open, the $Ar^+$ gun is differentially pumped by a turbo molecular pump, the filament is heated and the high power supply is switched off. The flow of neutral Ar gas from the ion gun into the main UHV system defines the pressure of the system, which is $< 1 * 10^{-8}$ mbar. The results are grouped into three categories 'instrument in operation for longer time', 'new monochromator & heating' and 'instrument open, arm repair & heating'.

The category 'instrument in operation for longer time' represents the normal operation mode of the XPS microprobe. The UHV was maintained for a longer time. In this operation mode samples mounted on sample holders were introduced via the turbo pumped intro chamber and the ion gun is operated for sample cleaning and sputter depth profiling, if necessary. So the only vacuum load for the UHV system comes from the samples themselves, the Ar gas and the residual gases of the intro chamber. Since the intro chamber is not heated, water is expected to be the significant load. In this operation mode small recontamination rates below ~ $1.5*10^{-3}$ [at% 'O' / sec] were estimated. Hence for a reactive material, which has high sticking coefficients for gas absorption, one should calculate with the build-up of a 1 at% recontamination within ~ 10 min. after sputtering. During two measurements the $Ar^+$ ion gun was operated without differential pumping. It gives data points at higher UHV system pressure. But even with this higher pressure the recontamination rate remains low. The higher impinge rate of Ar atoms on the instruments inner surfaces did not increase the contamination rate significantly.

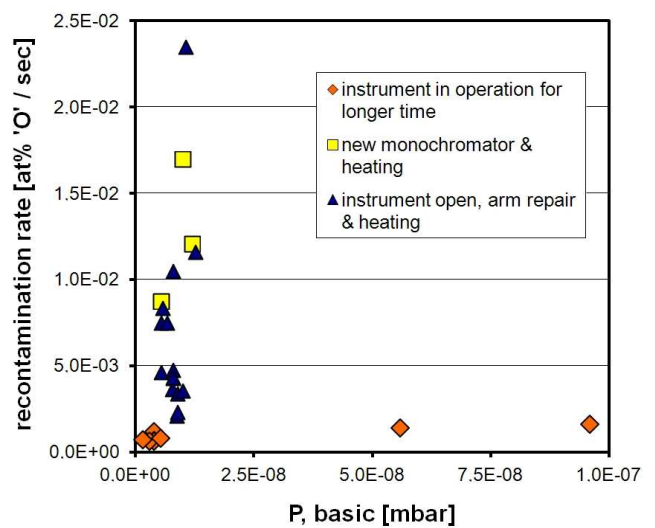

Fig. 3: Ti recontamination rate as function of UHV pressure

The category 'new monochromator & heating' stands for a system upgrade. The 'instrument open, arm repair & heating' represents an extensive repair of the sample handling system. In both cases many new mechanical components were mounted inside the UHV. Even though the pressure of the UHV system is below ~ $1*10^{-8}$ mbar recontamination rates up to $2.3*10^{-2}$ [at% 'O' / sec] were estimated. This demonstrates that the recontamination rate is the better measure for the instruments sputter depth profiling quality than the system pressure.

Fig. 4 shows the development of the Ti recontamination rate after the repair of the sample handling system as function of time. It needs ~ 20 days to come back to a low Ti recontamination rate even so the UHV pressure with the ion gun in standby was below ~ $8*10^{-9}$ mbar all the time. The single data point measured after 45 days shows, that the precious measurement of outgasing samples influences the next Ti recontamination rate measurement. Generally, the recontamination rate first decreases over time and then it converge against a limit of ~ $3*10^{-3}$ [at% 'O' / sec].

## 4. Conclusions

Obviously the low pressure of the UHV system of an analytical sputter depth profiling instrument is an obligatory prerequisite. But as demonstrated here for a Quantum 2000 XPS microprobe, there exists no direct correlation between a low sample recontamination rate after sputtering of the sample surface and the pressure in the UHV chamber.

As follows the new method developed to monitor the surface recontamination behavior of a sputter depth profiling instrument: A Ti sample is sputtered inside the instrument and the O recontamination of this surface is measured as a function of time. At the beginning the O surface contamination in-creases linear with time. The estimated recontamination rate describes the actual performance of the instrument. The data can be recorded within every sputter depth profiling instrument. Obviously every instrument has the necessary hardware. Only the pure Ti sample has to be supplied. This method can be used for a performance monitoring of every analytical sputter depth profiling instrument. It determines the instruments performance using exactly those instrumental components which are used for sputter depth profiling.

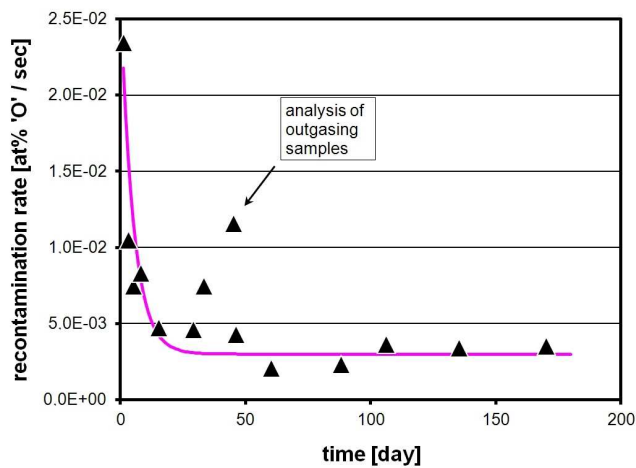

Fig. 4: development of the Ti recontamination rate after the repair of the sample handling system

In a worst-case scenario the recontamination rate of the instrument allows to estimate the time needed to build up a certain contamination level via adsorption of residual gas particles. For the XPS microprobe evaluated here, on a high sticking coefficient material a 1 at% recontamination will build–up within approximately 10 minutes. This adsorption process competes with the time needed for the measurement of element specific signals with the intended detection limit. In special cases the sample surface recontamination can impede to measure elements, which are sample components, with the desired detection limits. But fortunately often the situation is less critical, because only a few materials are expected to have such high sticking coefficients as Ti has.

## Acknowledgement

All measurements were done utilizing the Quantum 2000, instrument no. 78, installed at Siemens AG, Munich, Germany. I acknowledge the permission of the Siemens AG to use the measurement results here. For fruitful discussions and suggestions I would like to express my thanks to my colleagues.